\documentclass[11pt, hidelinks, a4paper]{scrartcl}
    \usepackage{amsmath,amssymb}
    \usepackage{booktabs}  
    \usepackage{authblk}  
    \usepackage{caption}
    \usepackage{comment}
    \usepackage[top=30truemm,bottom=35truemm,left=30truemm,right=30truemm]{geometry}
    \usepackage{graphicx}
    \usepackage{hyperref}
    \usepackage{multicol}
    \usepackage[section]{placeins} 
    \usepackage{xspace}

    \newcommand{\tn}[1]{\textnormal{#1}}
    \newcommand{\eV}{\text{e\kern-0.15ex V}\xspace}
    \newcommand{\GeV}{\text{G\eV}\xspace}
\graphicspath{{img_proceedings/}}

\title{\textbf{A combined fit to the Higgs Branching Ratios at ILD} }
\author{Jonas~Kunath\thanks{Presenter.
    Talk presented at the International Workshop on Future Linear Colliders (LCWS2021),
    15-18 March 2021. C21-03-15.1}
}
\author{Fabricio~Jimenez~Morales}
\author{Jean-Claude~Brient}
\author{Vincent~Boudry}
\affil{Laboratoire Leprince-Ringuet CNRS, \'Ecole Polytechnique,\hspace{2em} Institut Polytechnique de Paris, France}
\date{}
\begin{document}
\maketitle
\begin{abstract}
Higgs decay branching ratios at future Higgs factories
can be measured by directly exploiting class numeration.
Given the clean environment at a lepton collider,
it is possible to build an event sample highly enriched in Higgs bosons
and essentially unbiased for any decay mode.
The sample can be partitioned into categories using event properties
linked to the expected Higgs decay modes.
The counts per category are used to fit
the Higgs branching ratios in a model independent way.
The result of the fit is the set of branching ratios,
independent from a Higgs production mode measurement.

We present a simplified study on simulated data
for the International Linear Detector (ILD)
at the International Linear Collider (ILC) at 250~\GeV center-of-mass energy.
\end{abstract}

\section{Introduction}
The ILC as a Higgs factory will produce a large number of Higgs bosons
in the comparatively simple collision environment of a lepton collider.
The projected gain on the precision of Higgs measurements
will make them sensitive to deviations
that are predicted by theories of physics beyond the Standard Model (BSM).

With the Higgsstrahlung process at a Higgs factory
it is possible to construct an unbiased sample
in which each type of Higgs decay occurs with the probability given by
its branching ratio (BR) realized in nature~\cite{HiggsBR_LCWS2019}.
Higgsstrahlung events with $Z \to e^+ e^-$ and $Z \to \mu^+ \mu^-$
can be selected by reconstruction of the primary Z boson,
independently of the Higgs decay \cite{hRecoilIndependence}.

\begin{figure}[ht]
    \centering
    \includegraphics[width=0.49\textwidth, keepaspectratio]{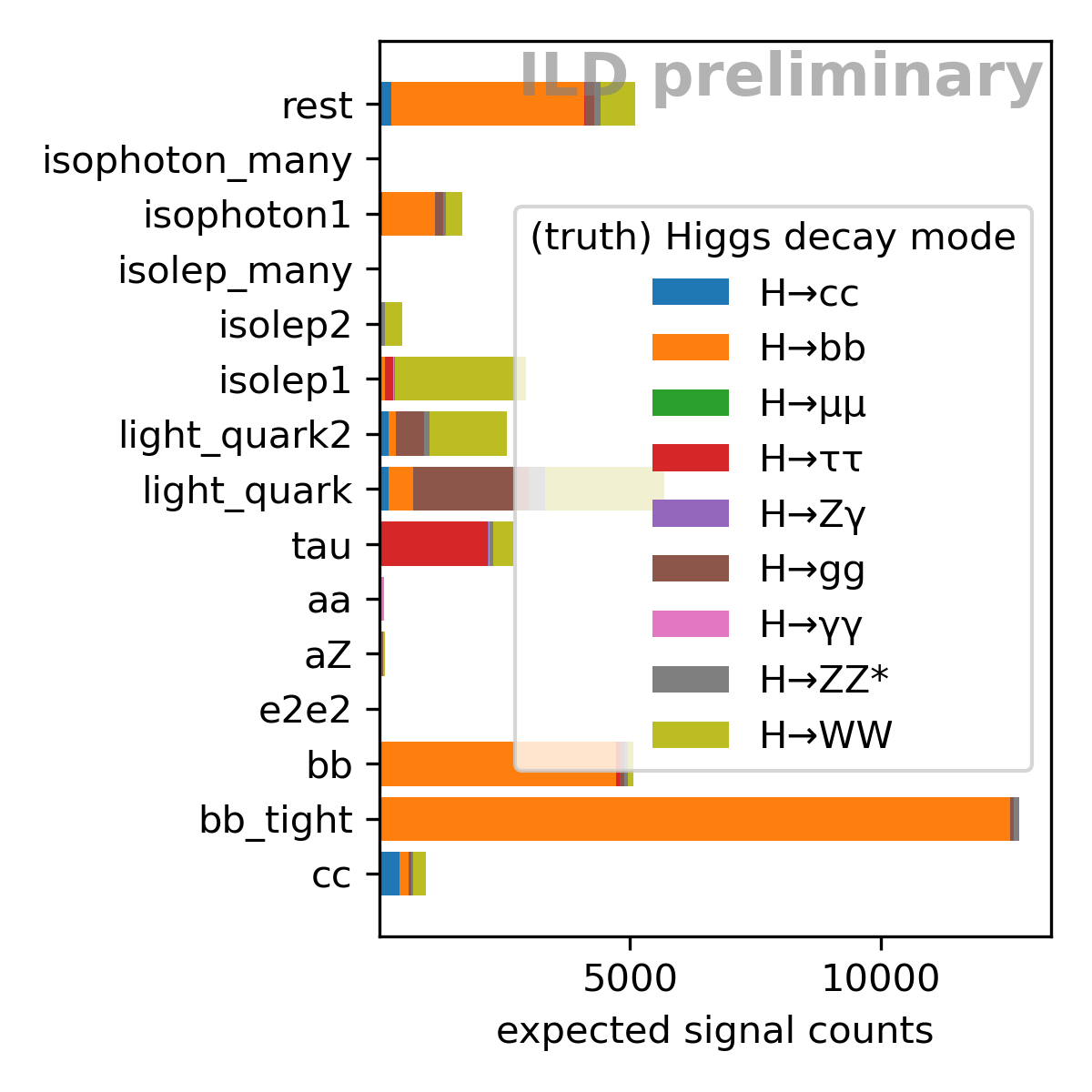}
    \caption{
        Expected contributions per class from each of the Higgs decay modes
        assuming the Standard Model branching ratios.
        The class definitions are listed in the appendix.
    }\label{fig:box_counts}
\end{figure}

Figure~\ref{fig:box_counts} shows how such a sample can be partitioned
into a number of classes with each of the class selection efficiencies depending
on the specific Higgs decay mode.
The relative frequency of the number of events per class
depends on the Higgs BRs.
An inclusive estimation of all branching ratios can be obtained
through a fit on the class counts.
The approach naturally lends itself to an extension with additional classes
which target BSM Higgs decays.
Additional decay modes can thus be excluded
with an upper limit that depends on the sample size.

The presented study is realized in the context of the
International Linear Detector (ILD)~\cite{ILD_DBD,ILD_IDR}.
The ILD detector concept is proposed for the International Linear Collider (ILC).
The detector design is based on the particle flow approach
for particle reconstruction~\cite{ParticleFlow}.
Only running scenarios of the ILC
with a center-of-mass energy of $\sqrt{s}=250~\GeV$ (ILC250)
are considered here.

The fit is in a preliminary state.
The simulation procedure is described in Section~\ref{sec:simulation},
a summary of the
current simplifications in~\ref{subsec:simplifications}.
Preliminary results from the fit,
with some known shortcomings, are already available.
They are given in Section~\ref{sec:fit}.
An outline of the necessary future work is presented
in Section~\ref{sec:future_work}.
Finally, the note is summarized in Section~\ref{sec:conclusion}.

\section{Simulation}\label{sec:simulation}
We use samples produced by the ILD concept group since 2020.
They are obtained from a detailed simulation of the ILC250
with the new SetA beam parameters~\cite{ILC_Staging_2017}.

The events are generated at leading order using
WHIZARD version 2.8.5~\cite{whizard,omega}.
Initial State Radiation, Beamstrahlung and Final State Radiation are included.
The fragmentation and hadronization of final-state quarks and gluons
is performed with PYTHIA 6.422~\cite{pythia}.
The ILD detector geometry is described with DD4hep~\cite{DD4hep}
and simulated in GEANT4~\cite{GEANT4}.
Event reconstruction is performed with ILCSoft v02-02~\cite{ILCSoft},
which includes PandoraPFA~\cite{PandoraPFA} for
the construction of particle flow objects
and LCFIPlus~\cite{LCFIPlus} for flavor tagging.

As the simulated data set is used for two different tasks,
it is split into two equal but statistically independent parts.
The first part of the available data (MC1) is needed
to set up the fit of BRs.
The second part (MC2) is used to generate the counts per class.
This second part serves as a placeholder for the detector data.
Having MC2 as a separate sample is required
for being able to evaluate the bias
that the method has due to the limited size of MC1.
By changing the BRs in MC2, through changing the event weights,
the adaptability of the fit is validated.
This is illustrated in Figure~\ref{fig:toys}.

\subsection{Current simplifications}\label{subsec:simplifications}
Results shown here are obtained without considering background processes.
As signal channel, $\nu \bar{\nu} H$ is used
without a pre-selection step for background reduction.
While it is not a realistic scenario,
this pure-Higgs sample can provide a first idea
of the precision and adaptability of the method.

For each of the nine considered main BRs of the Higgs boson,
at least 400k simulated events are used.
They are separately weighted and combined into a single sample
with exactly 40k signal Higgs events.
Different Higgs BR scenarios are tested by changing the weights.

Note that 40k Higgs events corresponds to about 10\% of the Higgs bosons
produced in the long-running H-20 scenario of the ILC250~\cite{ILC_Scenarios}.
Even before taking into account selection inefficiencies,
only less than 7\% of the produced Higgs bosons
are in the $Z \to e^+ e^-$ and $Z \to \mu^+ \mu^-$ samples combined.
As mentioned in Section \ref{sec:future_work}
we expect to be able to leverage the $\nu \bar{\nu} H$ sample.
Future work will have to show if a 10\% sample size
is a good estimation of the gain from the $\nu \bar{\nu} H$ sample.

\section{Branching ratio uncertainties}\label{sec:fit}

For each Higgs decay mode, a probability vector is
constructed from the simulation (MC1).
This vector stores the selection efficiency and the probabilities
for the events of the chosen type to belong to each of the classes
that are constructed within the sample.
The vectors can be collected as columns into a probability matrix $M$.
Data counts are generated from the test sample (MC2)
as placeholders for future detector data.

A fit is then performed with \texttt{iminuit}~\cite{Minuit,iminuit}.
The Standard Model (SM) Higgs branching ratios (BRs)
are taken
as the starting values of the fit.
Based on a multinomial log-likelihood as its cost function,
the 9 considered independent BRs of the Higgs boson
are described through 8 parameters in the fit.
We obtain a prediction for the 8 parameters
and the corresponding covariance matrix.
From those we get the mean values and uncertainties of the BRs.

\begin{figure}[ht]
    \centering
    \includegraphics[width=0.49\textwidth, keepaspectratio]{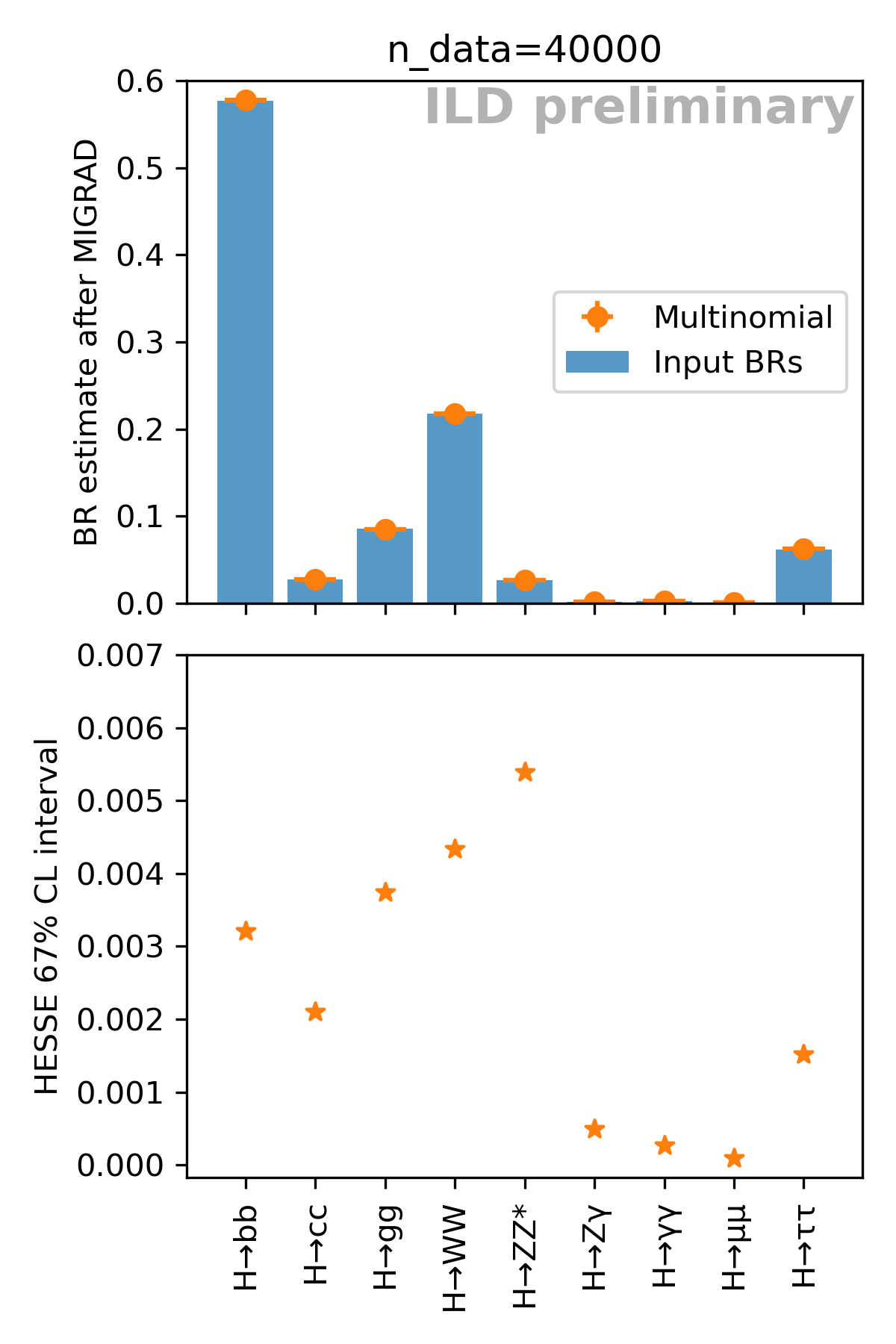}
    \includegraphics[width=0.49\textwidth, keepaspectratio]{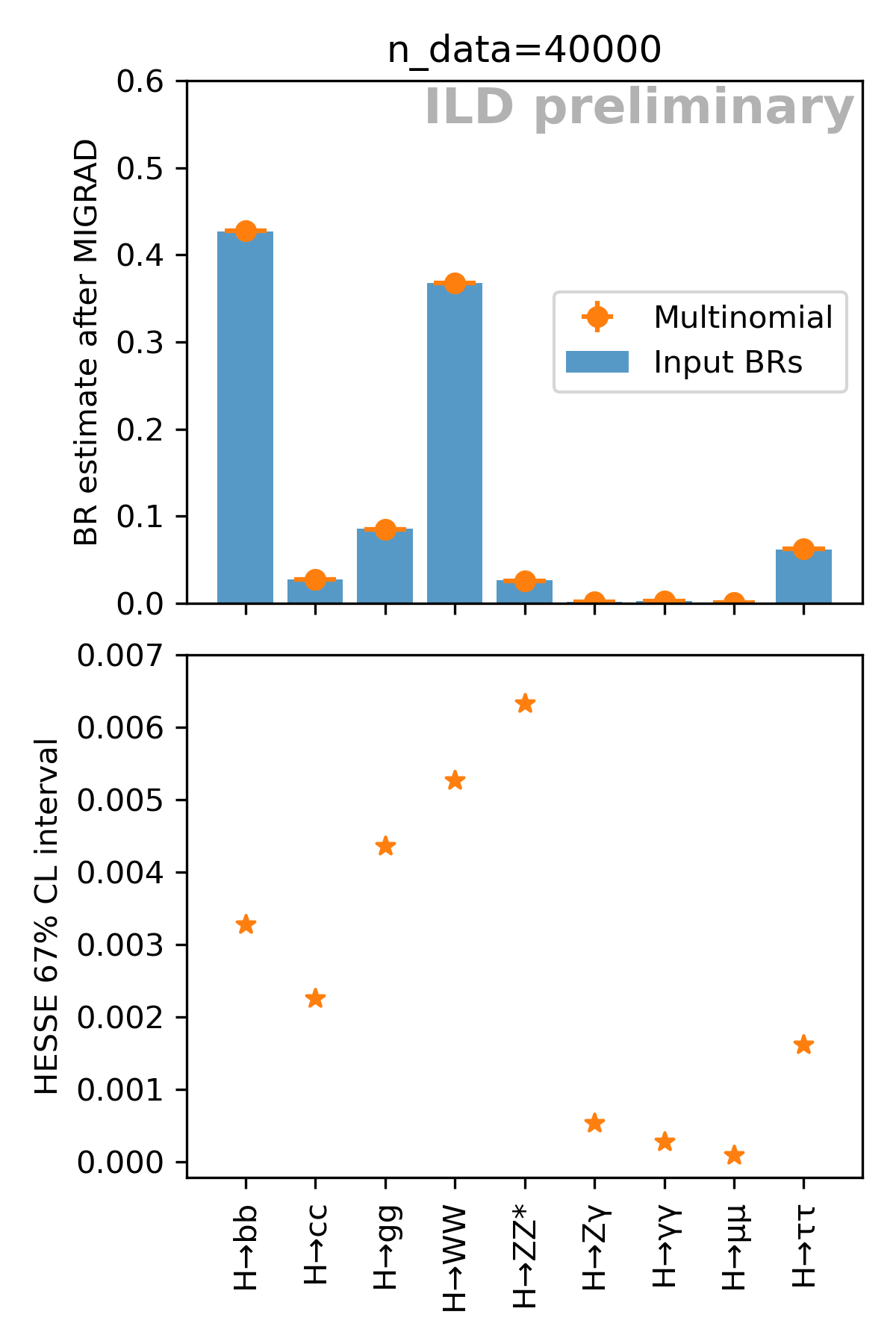}
    \caption{
        Top row: Branching ratios assumed for the generation of the counts per class.
        Fit optimum and its uncertainty are given as orange error bars.
        The starting values of the fit are always the SM Higgs BRs.
        The considered scenarios are the SM BRs (left)
        and a BSM scenario described in the text.
        \\
        Bottom row: The absolute statistical uncertainties
        of the fit per branching ratio.
    }\label{fig:brs}
\end{figure}

The result of this fit is shown in Figure~\ref{fig:brs} for
the SM BRs and an alternative scenario.
To have a change that can be comfortably spotted by eye,
the alternative scenario has the $BR(H \to b \bar{b})$ decreased by $15\%$,
which is fully compensated by $BR(H \to W^+W^-)$.
Since the probability vectors of the decay modes are different enough,
the fit moves to the altered BRs without any problems.
The fit not landing at the altered BRs
would have suggested that the defined classes
do not have sufficient discriminating power
between some of the BRs.

\begin{figure}[ht]
    \centering
    \includegraphics[width=0.49\textwidth, keepaspectratio]{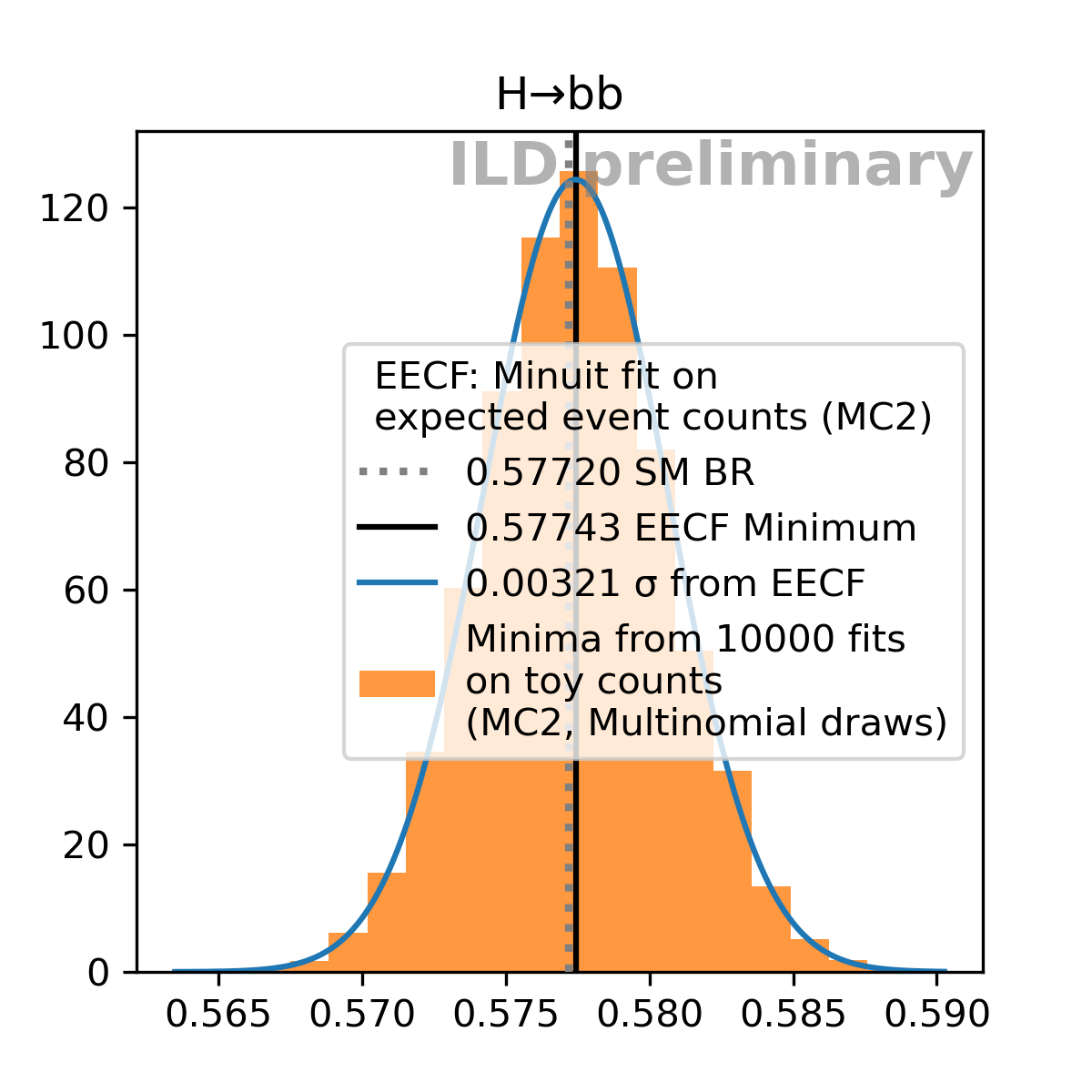}
    \caption{
        Validation of the uncertainty from
        a fit on the expected event counts scenario against a toy study.
    }\label{fig:toys}
\end{figure}

The uncertainties on the BRs can be tested in a toy study.
The uncertainties and correlations of the fit on the expected event counts
are obtained through the second derivatives of the likelihood function
at the fit optimum.
For the toy study, samples are drawn from a multinomial distribution centered on the
expected counts for each class.
Then, a fit is performed for each of the samples, and the predicted
branching ratio optima are collected.
An example toy study on the $H \to b \bar{b}$ branching ratio is shown in
Figure~\ref{fig:toys}.
The dotted grey line indicates the SM value of
the $H \to b \bar{b}$ coupling in the simulation.
The black line indicates the optimum of a fit on
the expected event counts.
It was validated that the difference between these two lines
is only due to limited Monte Carlo statistics.
The blue Gaussian curve has as standard deviation the uncertainty of
the fit on the expected event counts, as quoted by \texttt{MINUIT}.
The orange histogram contains the fit optima from
10k fits on event counts varied according to
a multinomial distribution centered on the expected event counts.
The values from the toy study agree well with those anticipated
from the fit of the expected event counts.

The projected (absolute) $1 \sigma$-uncertainties for the Higgs branching ratios
are listed in Table~\ref{tab:br_uncertainties}
and illustrated in Figure~\ref{fig:brs} (bottom row).
The numbers are based on the simplified scenario
outlined in Section~\ref{subsec:simplifications}.

\begin{table}[ht]
    \centering
    \begin{tabular}{lrrr}
        \toprule
        {} &  SM BR &  fitted BR &  $\sigma_{\tn{stat}}$ \\
        \midrule
        $H\to cc$           &  2.718 &    2.733 &     0.210 \\
        $H\to bb$           & 57.720 &   57.743 &     0.321 \\
        $H\to \mu\mu$       &  0.030 &    0.030 &     0.009 \\
        $H\to \tau\tau$     &  6.198 &    6.207 &     0.152 \\
        $H\to Z\gamma$      &  0.170 &    0.176 &     0.050 \\
        $H\to gg$           &  8.550 &    8.499 &     0.374 \\
        $H\to \gamma\gamma$ &  0.242 &    0.243 &     0.027 \\
        $H\to WW$           & 21.756 &   21.761 &     0.434 \\
        $H\to ZZ^*$         &  2.616 &    2.608 &     0.539 \\
        \bottomrule
        \end{tabular}
    \caption{
        Preliminary results of a \texttt{MINUIT} fit on the expected event counts.
        The table gives fitted values of the Higgs BRs
        and their absolute statistical uncertainties.
        All numbers are given in percent. \\
        }\label{tab:br_uncertainties}
\end{table}

\section{Future work}\label{sec:future_work}
The next iteration of this study has to include the background rates for
Higgs-free events passing the selection.
Then, the proper cross sections and the polarizations foreseen at ILC
can be applied.

In the $\nu \bar{\nu} H$ sample it is not possible to
perform an event selection which is independent of the Higgs decay products.
Thus, this sample alone cannot be used for a branching ratio derivation with
an unconditional upper limit on the unobserved branching ratio.
But in combination with the other samples, the larger statistics of this
sample should help decrease the uncertainties.
The $\nu \bar{\nu} H$ sample should be added to the combined fit,
after choosing a good pre-selection.

The decay products from the  $Z \to e^+ e^-$ and $Z \to \mu^+ \mu^-$ decays
of the primary Z boson in a Higgsstrahlung event can be reliably identified.
They can be separated from the rest of the event
and exploited for the event selection.
Since the identification of all particles from
$Z \to \tau^+ \tau^-$ or $Z \to q \bar{q}$ decays
can be more complicated
it is not planned to use those Higgsstrahlung events.

The current class definitions are listed in the appendix.
They are only an initial draft with potential for improvement.
The rest class should be reserved almost exclusively for
background (or unexpected Higgs decay) events.
For some of the Higgs decays, including $H \to W^+ W^-$,
no proper classes are designed yet.

A mechanism that indicates the case of non-SM Higgs decays
($H \to \mu^+\tau^-, H \to b\bar{c}$, \ldots)
has to be established.
This could be through one or more additional classes
or through a comparison of the value of the likelihood at the fit minimum
with the likelihood minimum for the expected class counts.

\section{Conclusion}\label{sec:conclusion}
An simultaneous measurement of all the Higgs branching ratios at once is
an appealing objective.
It can give upper limits for unobserved Higgs decays.
The measurements of the individual branching ratios are
connected through the covariance matrix of the fit.
The preliminary results are attractive and motivate
improvements towards a more realistic analysis.

\section*{Acknowledgements}
The authors would like to thank the LCC generator working group and the ILD software working
group for providing the simulation and reconstruction tools and producing the Monte Carlo
samples used in this study. This work has benefited from computing services provided by
the ILC Virtual Organization, supported by the national resource providers of the EGI
Federation and the Open Science GRID.

\newpage 
\providecommand{\href}[2]{#2}\begingroup\raggedright
\endgroup 

  \newpage
  \appendix
  \section*{Appendix}
  \subsection*{Class definitions}
  The current class definitions are only an initial attempt.
  They will be refined and extended.

  An event belongs to the first class for which it passes the selection.
  The class selections are applied in the order listed below.
  It is not necessary that a class has a high purity for a specific branching ratio.
  The relative contributions that are expected from the Higgs decay modes
  for each of the classes
  should be disparate.

  All variables are built exclusively on the Higgs-candidate part of the event
  excluding isolated leptons and final state radiation photons from the primary Z boson decay.
  Isolated leptons and isolated photons are identified through the
  IsolatedLeptonTaggingProcessor
  and IsolatedPhotonTaggingProcessor~\cite{ILCSoft}.

  The flavor tagging is performed after forcing the (Higgs part of the)
  event into two jets.
  Through LCFIPlus~\cite{LCFIPlus} we obtain a score for b-likeliness ($btag$)
  and for c-likeliness ($ctag$) for each of the two jets.
  Each score lies between 0 and 1.
  Their sum cannot exceed 1.
  The sum is small when the jet is identified as likely stemming from a light quark.

  {\small\begin{multicols}{2}
  \begin{enumerate}
      \item \texttt{cc}: Targets $H \to c \bar{c}$:
      \begin{itemize}
          \item No isolated leptons or photons.
          \item $M_H > 100~\GeV$.
          \item More than 20 Particle Flow Objects (PFOs).
          \item $ctag1 > 0.5$, $ctag2 > 0.5$.
      \end{itemize}
      \item \texttt{bb\_tight}: Targets $H \to b \bar{b}$:
      \begin{itemize}
          \item No isolated leptons or photons.
          \item $btag1 > 0.8$, $btag2 > 0.8$.
      \end{itemize}
      \item \texttt{bb}: Also targets $H \to b \bar{b}$:
      \begin{itemize}
          \item No isolated leptons or photons.
          \item $btag1 > 0.8$.
      \end{itemize}
      \item \texttt{e2e2}: Targets $H \to \mu^+ \mu^-$:
      \begin{itemize}
          \item Has an opposite-charge pair of isolated muons.
          \item $M_{\mu^+ \mu^-} \in \left[100~\GeV, 130~\GeV \right]$.
      \end{itemize}
      \item \texttt{aZ}: Targets $H \to \gamma Z$:
      \begin{itemize}
          \item Has an isolated photon.
          \item $M_{\gamma} \in \left[20~\GeV, 50~\GeV \right]$.
          \item $\left| \tn{cos}\theta_{\gamma} \right| < 0.9$.
          \item $M_{\gamma} \in \left[75~\GeV, 100~\GeV \right]$,
              where Z is everything but the photon.
      \end{itemize}
      \item \texttt{aa}: Targets $H \to \gamma \gamma$:
      \begin{itemize}
          \item Has an isolated photon and no isolated leptons.
          \item Less than 15 PFOs.
          \item $E_H > 125~\GeV$.
          \item $\left| \tn{cos}\theta_{\gamma} \right| < 0.9$.
          \item $E_{\gamma} > 35~\GeV$.
      \end{itemize}
      \item \texttt{tau}: Targets $H \to \tau^+ \tau^-$:
      \begin{itemize}
          \item Has no isolated leptons.
          \item Less than 15 PFOs.
      \end{itemize}
      \item \texttt{light\_quark}: Targets $H \to gg$:
      \begin{itemize}
          \item Has no isolated leptons or photons.
          \item $btag1 + ctag1 < 0.5$.
      \end{itemize}
      \item \texttt{light\_quark2}: Also targets $H \to gg$:
      \begin{itemize}
          \item Has no isolated leptons or photons.
          \item $btag2 + ctag2 < 0.5$.
      \end{itemize}
      \item \texttt{isolep1}: Targets $H \to W^+ W^-$, $H \to \tau^+ \tau^-$:
      \begin{itemize}
          \item Exactly 1 isolated lepton.
      \end{itemize}
      \item \texttt{isolep2}: Targets $H \to W^+ W^-$, $H \to ZZ^*$:
      \begin{itemize}
          \item Exactly 2 isolated leptons.
      \end{itemize}
      \item \texttt{isolep2}: Targets $H \to ZZ^*$:
      \begin{itemize}
          \item More than 2 isolated leptons.
      \end{itemize}
      \item \texttt{isophoton1}: No specific target:
      \begin{itemize}
          \item Exactly 1 isolated photon.
      \end{itemize}
      \item \texttt{isophoton\_many}: No specific target:
      \begin{itemize}
          \item More than 1 isolated photon.
      \end{itemize}
      \item \texttt{rest}: No specific target:
      \begin{itemize}
          \item Takes whatever did not belong to any of the previous classes.
      \end{itemize}
  \end{enumerate}
\end{multicols}}
\end{document}